\documentclass[conference]{IEEEtran}
\IEEEoverridecommandlockouts
\usepackage{cite}
\usepackage{amsmath,amssymb,amsfonts}
\usepackage{algorithmic}
\usepackage{graphicx}
\usepackage{textcomp}
\usepackage{xcolor}
\usepackage{mathtools}
\usepackage{booktabs}
\usepackage{enumitem}
\usepackage{xurl}
\def\BibTeX{{\rm B\kern-.05em{\sc i\kern-.025em b}\kern-.08em
    T\kern-.1667em\lower.7ex\hbox{E}\kern-.125emX}}
\begin{document}

\title{Pay (Cross) Attention to the Melody: Curriculum Masking for Single-Encoder Melodic Harmonization\\
% \thanks{Acknowledgments and Disclosure of Funding
% This work has been partially supported by project MIS 5154714 of the National Recovery and
% Resilience Plan Greece 2.0 funded by the European Union under the NextGenerationEU Program.}
}

\makeatletter
\newcommand{\linebreakand}{%
  \end{@IEEEauthorhalign}
  \hfill\mbox{}\par
  \mbox{}\hfill\begin{@IEEEauthorhalign}
}
\makeatother

\author{

\IEEEauthorblockN{Maximos Kaliakatsos-Papakostas}
\IEEEauthorblockA{
\textit{Department of Music Technology and Acoustics} \\
\textit{Hellenic Mediterranean University, Greece} \\
\textit{Archimedes, Athena RC, Greece}
}

\and

\IEEEauthorblockN{Dimos Makris}
\IEEEauthorblockA{
\textit{Department of Music Technology and Acoustics} \\
\textit{Hellenic Mediterranean University, Greece} \\
\textit{Archimedes, Athena RC, Greece} \\
\textit{~}
}

\linebreakand

\IEEEauthorblockN{Konstantinos Soiledis}
\IEEEauthorblockA{
\textit{Department of Music Technology and Acoustics} \\
\textit{Hellenic Mediterranean University, Greece} \\
\textit{Archimedes, Athena RC, Greece} \\
\textit{~}
}

\and

\IEEEauthorblockN{Konstantinos-Theodoros Tsamis}
\IEEEauthorblockA{
\textit{Department of Music Technology and Acoustics} \\
\textit{Hellenic Mediterranean University, Greece} \\
\textit{Archimedes, Athena RC, Greece} \\
\textit{~}
}

\linebreakand

\IEEEauthorblockN{Vassilis Katsouros}
\IEEEauthorblockA{
\textit{Institute of Language and Speech Processing} \\
\textit{Archimedes, Athena RC, Greece} \\
\textit{~}
}

\and

\IEEEauthorblockN{Emilios Cambouropoulos}
\IEEEauthorblockA{
\textit{School of Music Studies} \\
\textit{Aristotle University of Thessaloniki, Greece} \\
\textit{~} \\
\textit{~}
}

}

\maketitle

\begin{abstract}

Melodic harmonization, the task of generating harmonic accompaniments for a given melody, remains a central challenge in computational music generation. Recent single encoder transformer approaches have framed harmonization as a masked sequence modeling problem, but existing training curricula inspired by discrete diffusion often result in weak (``cross'') attention between melody and harmony. This leads to limited exploitation of melodic cues, particularly in out-of-domain contexts. In this work, we introduce a training curriculum, \texttt{FF} (full-to-full), which keeps all harmony tokens masked for several training steps before progressively unmasking entire sequences during training to strengthen melody–harmony interactions. We systematically evaluate this approach against prior curricula across multiple experimental axes, including temporal quantization (quarter vs. sixteenth note), bar-level vs. time-signature conditioning, melody representation (full range vs. pitch class), and inference-time unmasking strategies. Models are trained on the HookTheory dataset and evaluated both in-domain and on a curated collection of jazz standards, using a comprehensive set of metrics that assess chord progression structure, harmony–melody alignment, and rhythmic coherence. Results demonstrate that the proposed \texttt{FF} curriculum consistently outperforms baselines in nearly all metrics, with particularly strong gains in out-of-domain evaluations where harmonic adaptability to novel melodic queues is crucial. We further find that quarter-note quantization, intertwining of bar tokens, and pitch-class melody representations are advantageous in the \texttt{FF} setting. Our findings highlight the importance of training curricula in enabling effective melody conditioning and suggest that full-to-full unmasking offers a robust strategy for single encoder harmonization.

\end{abstract}


\section{Introduction}\label{sec:intro}

Transformer architectures have emerged as powerful sequence modeling frameworks across domains such as language, vision, and music~\cite{vaswani2017attention, huang2018musictransformer}. Within symbolic music generation, melodic harmonization is a particularly challenging task: given a melodic sequence, the goal is to produce a harmonic sequence that is both locally compatible and globally coherent. This requires alignment of chords with melodic material while simultaneously maintaining harmonic progression structure over longer spans. As such, harmonization provides a rich testbed for exploring how sequence models integrate cross-modal signals (melody and harmony) across both local and global contexts.

Early neural approaches to melodic harmonization employed bidirectional LSTMs~\cite{lim2017chord, yeh2021automatic, chen2021surprisenet, costa2023neural}, while recent work has shifted to transformer-based models~\cite{huang2018musictransformer, rhyu2022translating, huang2024emotion, wu2024melodyt5}. These methods typically frame harmonization as a translation or summarization problem, where the melody is ``translated'' into a harmonic sequence that abstracts its structure. Most methods rely on autoregressive decoding, generating chords sequentially from left to right. Such models assume harmonic the dependency of new harmony tokens only on previous ones, which does not allow insertion of chord constraints prior to generation. Melodic harmonization is rather a bidirectional process, that occasionally involves setting harmonic checkpoints and then filling the blanks.

In parallel, diffusion models have gained traction for symbolic music generation, following developments in vision~\cite{ho2020denoising}. Some approaches operate in continuous symbolic spaces~\cite{mittal2021symbolic, lv2023getmusic}, others in latent VAE spaces~\cite{zhang2023fast}, and others on pianoroll images using U-Net backbones~\cite{atassi2023generating, min2023polyffusion, li2023melodydiffusion, wang2024whole, huang2024symbolic, zhang2025mamba}. While diffusion has not yet been applied directly to melodic harmonization, related work in text generation has shown the effectiveness of discrete denoising and unmasking strategies. MaskGIT~\cite{chang2022maskgit}, D3PMs~\cite{austin2021structured}, and other discrete diffusion methods~\cite{plasser2023discrete} iteratively refine masked token sequences, offering flexible conditioning and faster generation compared to autoregressive models. In symbolic music, hybrid transformer–diffusion models have been explored, either by applying diffusion to transformer logits~\cite{jonason2024symplex} or to latent codes learned through quantized autoencoders~\cite{zhang2024composer}. These approaches highlight the potential of token-based unmasking as a generative mechanism, with particular advantages for tasks requiring global conditioning.

Curriculum masking strategies are rooted in earlier denoising autoencoder research~\cite{vincent2008extracting, bengio2013generalized}, which showed that stronger corruption early in training forces models to learn robust representations rather than trivial identity mappings. In NLP, non-autoregressive transformers~\cite{ghazvininejad2019mask, wang2019bert} and the Levenshtein Transformer~\cite{gu2019levenshtein} adopted similar iterative unmasking strategies, gradually refining predictions from coarse to fine. Diffusion models formalized these ideas with carefully designed noise schedules~\cite{austin2021structured, nichol2021improved}, where exponential or cosine schedules were found to balance context availability with reconstruction difficulty. In multimodal contexts, regularization techniques have been proposed to prevent models from ignoring cross-modal signals~\cite{tsai2019multimodal, kim2021vilt}. Together, these findings suggest that curriculum design is crucial for ensuring models exploit conditioning information effectively, rather than defaulting to modality-specific self-attention.

Recently proposed approaches to melodic harmonization adopt single-encoder generative architectures, where melody is clamped in the beginning positions of the encoder and harmony tokens are progressively unmasked in the latter positions. Even though single-encoder architectures involve a single attention map for each head, subsets of weights play the role of ``self'' (harmony attending to harmony) and ``cross'' (harmony attending to melody) attention. Training methods of such architectures have been proposed that are inspired by discrete diffusion~\cite{kaliakatsos2025diffusion} for iteratively and bidirectionally unmasking harmony tokens given a melodic sequence. Such training methods, however, exhibit weak melody-to-harmony attention, with models often over-relying on ``self-attention'', and underutilizing melodic cues (``cross-attention''). To address this, we propose a more robust training curriculum, \texttt{FF} (full-to-full), that begins with fully masked harmonic sequences for several initial training steps and then progressively unmasks until only a single mask remains. This strategy encourages strong ``cross-modal'' (between melody and harmony) conditioning early in training, while gradually shifting toward fine-grained reconstruction. We further explore task-specific factors that may influence harmonization quality, including temporal resolution (quarter vs. sixteenth notes), structure integration (bar tokens vs. time signatures), and melody representation (pitch class vs. full range).

Our contributions are threefold: (1) we describe the problem of weak melody ``cross-attention'' in single-encoder generative transformers; (2) we demonstrate that the \texttt{FF} curriculum significantly improves melody–harmony integration, yielding state-of-the-art results in both in-domain and out-of-domain evaluations; (3) we provide empirical analysis across datasets, representations, and inference strategies, answering five core research questions about harmonization modeling. The code for the models, the training and the experiments can be found online~\footnote{\url{https://github.com/NeuraLLMuse/SingleEncoderMHCrossAttention.git}}. This page provides links to a demo page and example MIDI files.

\section{Method}\label{sec:method}

Single-encoder generation is performed through an iterative unmasking of tokens, given some visible (already unmasked) context. Unlike left-to-right autoregressive decoding, this formulation leverages bidirectional context, which is particularly important for melodic harmonization. Human composers typically do not harmonize strictly linearly through time; instead, specific melody passages may strongly constrain the appropriate chord choice, and once these are fixed, the surrounding harmony can be refined in light of the emerging global context. Such a paradigm is closely related to non-autoregressive ``mask-predict'' strategies in text generation~\cite{ghazvininejad2019mask}, and to inpainting approaches in symbolic music generation~\cite{pati2019learning}. Beyond modeling convenience, this generative process naturally accommodates user constraints: specific chords can be fixed at arbitrary positions before generation, enabling interactive human–AI collaboration.

Figure~\ref{fig:overview} illustrates an abstract overview of a single-encoder harmonization model. During inference, the model receives melody input in the first half of the encoder and masked harmony tokens in the second half; the task is to iteratively ``unmask'' the harmony positions. During training, subsets of harmony tokens are masked, and the model learns to recover them, following iterative refinement strategies similar to the one used in Mask-Predict~\cite{ghazvininejad2019mask}. A recent study introduced two such inference–training couplings for harmonization~\cite{kaliakatsos2025diffusion}. While both outperformed autoregressive baselines, close inspection revealed that melodic nuances were often underutilized. We analyze this limitation abstractly in the following paragraphs before detailing our proposed representation, architecture, and curriculum solution.

\begin{figure}[!ht]
\includegraphics[width=0.47\textwidth]{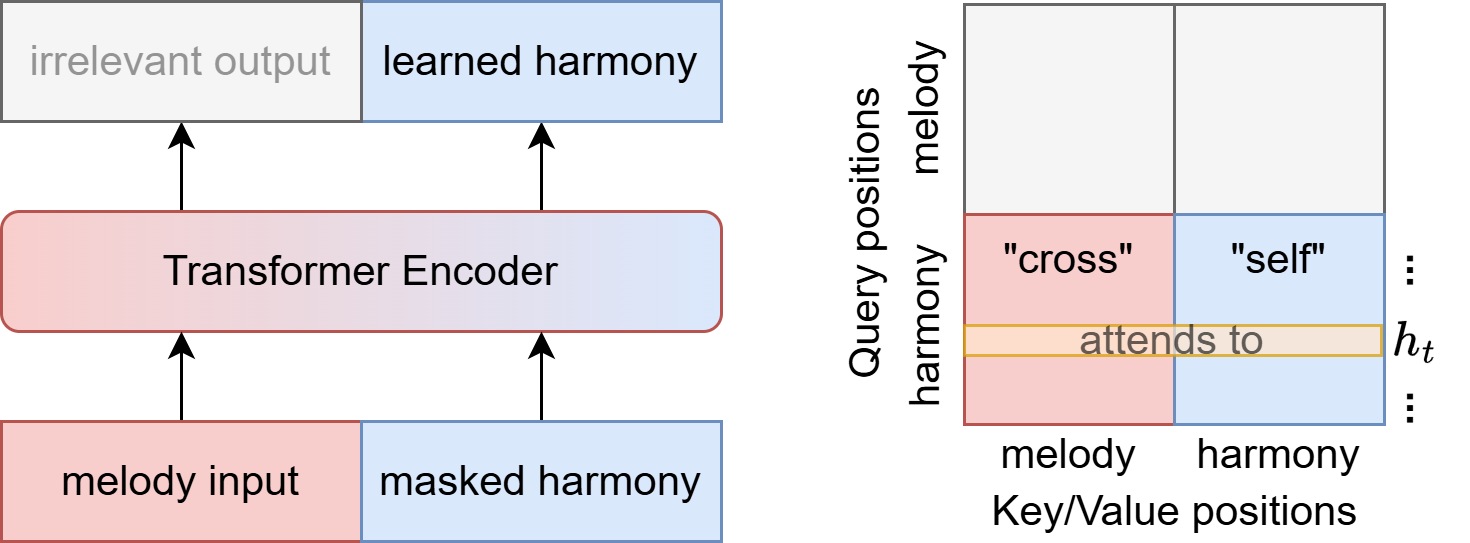}
\caption{Overview of the single-encoder architecture for melodic harmonization. Left: abstract representation of the transformer encoder input (melody in the first half, harmony in the second) and its output. Right: schematic of the encoder’s attention map. When predicting a harmony token at time step $t$ ($h_t$), only the lower half of the map is functionally relevant: the lower-left quadrant captures harmony–given–melody interactions (``cross'' attention), while the lower-right quadrant captures harmony–given–harmony interactions (``self'' attention).}\label{fig:overview}
\end{figure}

\subsection{Problem description}\label{subsec:problem}

In preliminary experiments with the two existing training and inference strategies (analyzed later), we observed that harmonically significant cues in the melody were often underutilized in the generated harmonizations. To better understand the nature of this issue, we constructed an artificial dataset where the correct harmonization could be determined solely from the melody. Specifically, the dataset consisted of 1,000 training and 100 test pieces, each 8 bars long with one chord per quarter note. Chords were chosen at random among the seven diatonic chords of C major, and melodies were built by placing the corresponding chord root as the melody note. By design, this procedure removed any exploitable chord-to-chord dependencies (no self-attention signal), while creating a one-to-one mapping between melody notes and chords (a strictly diagonal cross-attention relation). Similar use of synthetic diagnostic datasets has been employed in NLP to probe model behavior~\cite{ettinger2020bertnotlessonsnew, sinha2019clutrr}.

After training on this dataset with the two existing strategies and our proposed method, we examined average attention maps over all heads and layers during inference on a held-out test set piece of the artificial dataset. Figure~\ref{fig:attn_maps} summarizes the results: panels (a) and (b) illustrate that the existing methods did not recover the expected diagonal cross-attention pattern, while panel (c) shows that the proposed method successfully captured it. Importantly, even in this simplified setting where melody alone determines the harmony, the baseline strategies failed to leverage melodic information. A more detailed discussion of why this occurs follows after the presentation of the proposed method.

\begin{figure*}[!ht]
   \centering
\begin{tabular}{ccc}
\includegraphics[width=0.3\textwidth]{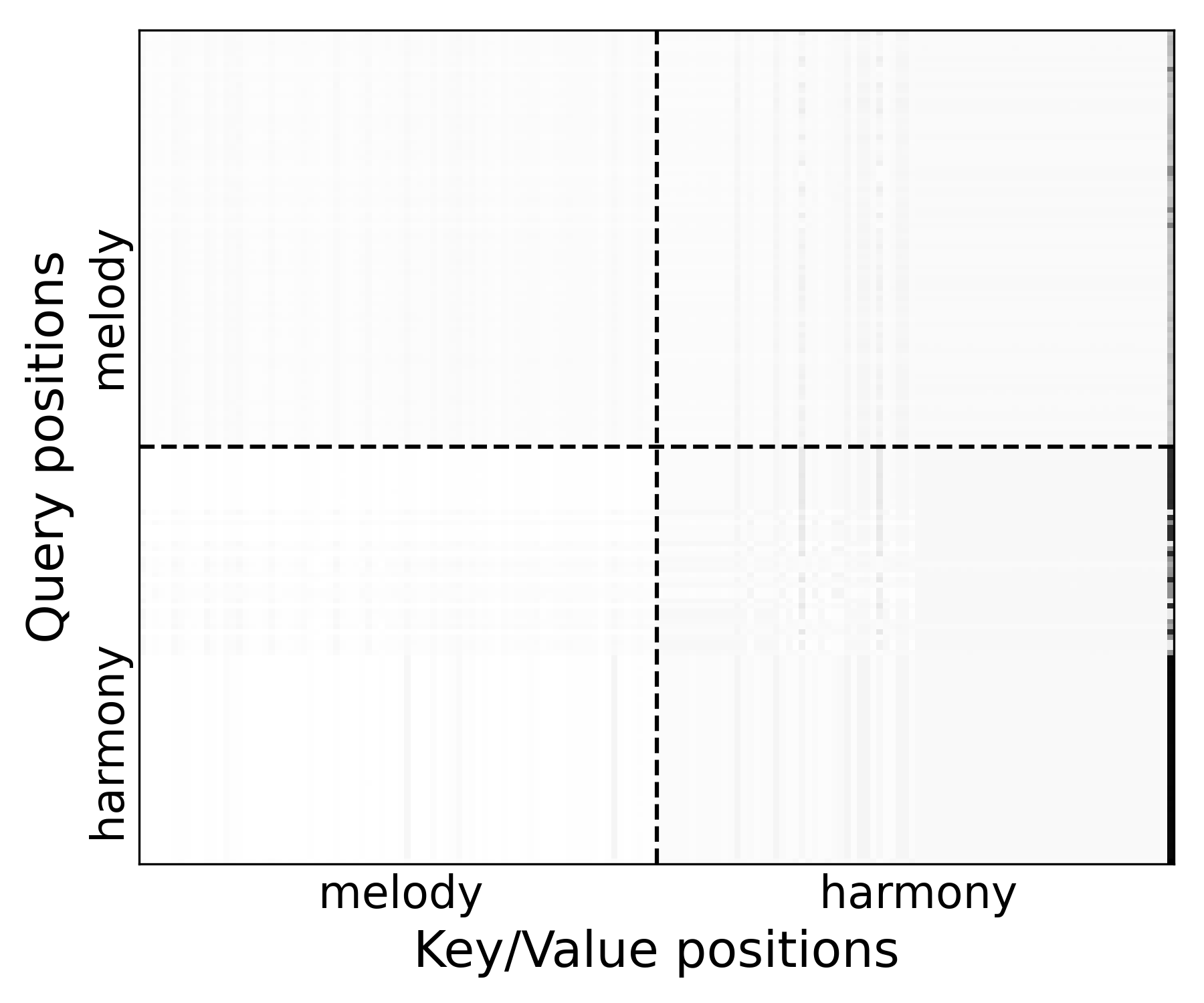}&
\includegraphics[width=0.3\textwidth]{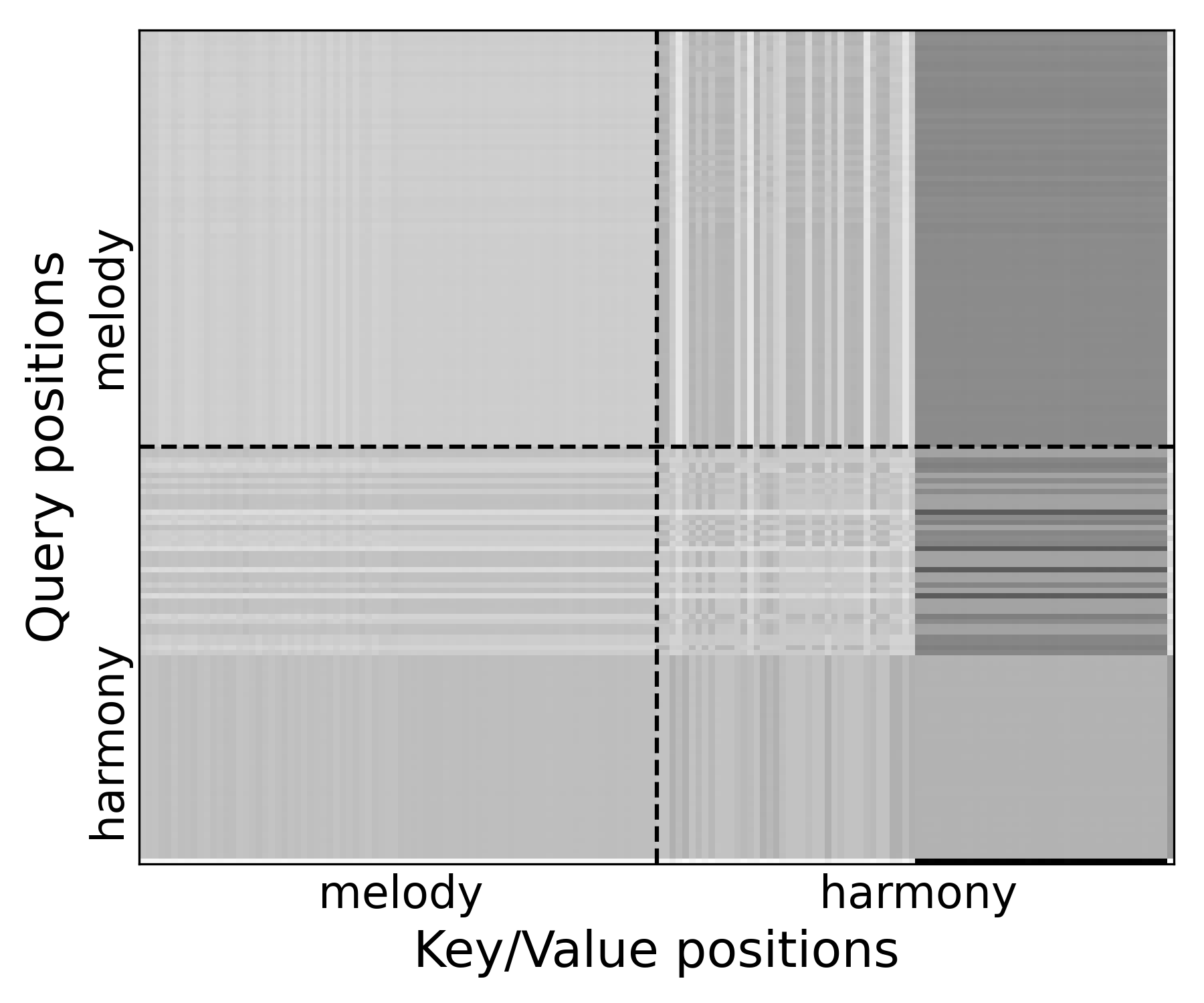}&
\includegraphics[width=0.3\textwidth]{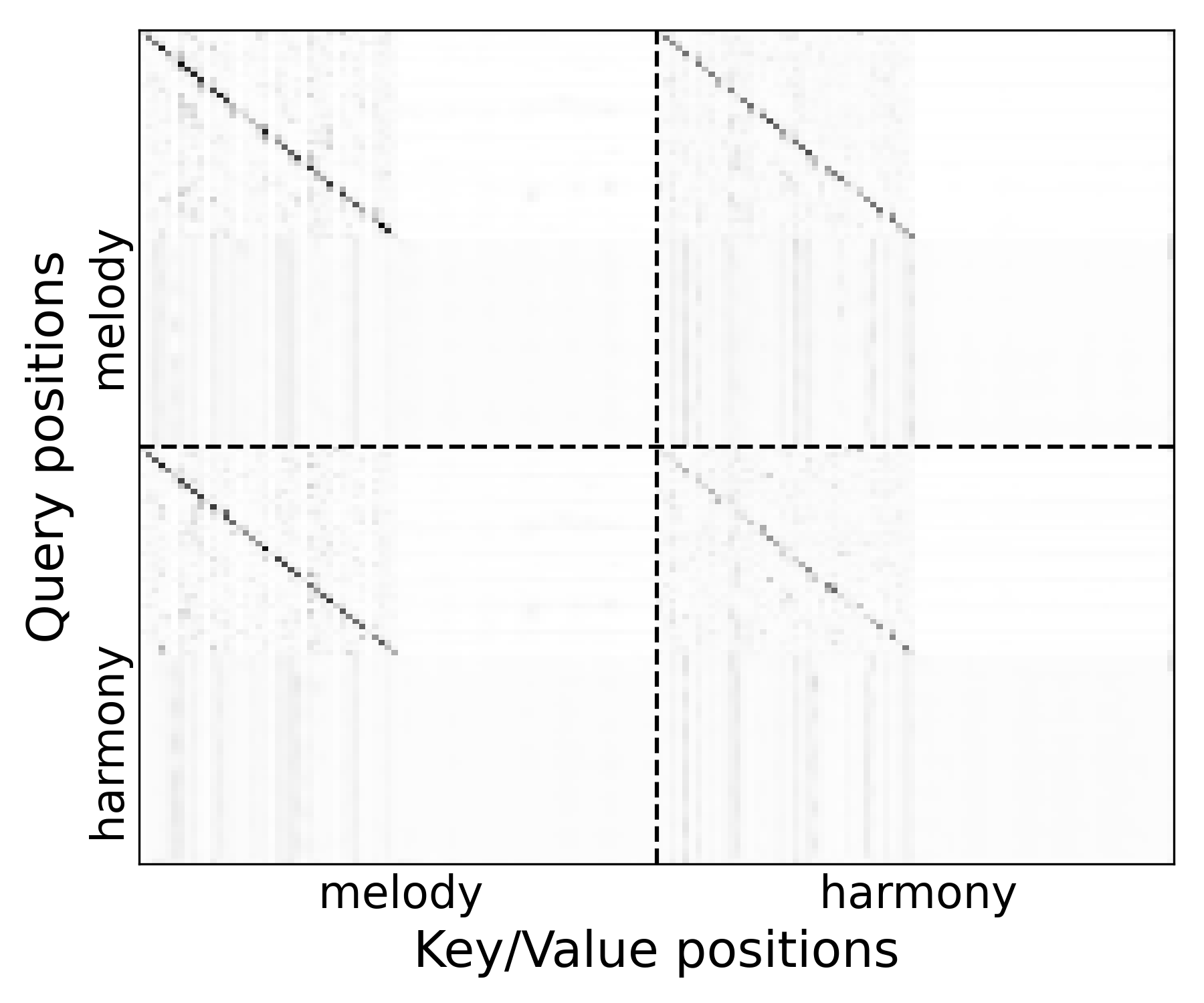}
\\
(a) Midpoint doubling (\texttt{MD}) & (b) Random 10\% (\texttt{R10}\%) & (c) Full-to-full (proposed \texttt{FF})
\end{tabular}
    \caption{Average attention maps across all layers and heads for the training methods from prior work (a, b) and for the proposed method (c). In (a) and (b), cross-attention from harmony to melody is largely absent, as indicated by the lack of diagonal patterns in the lower-left quadrant. In contrast, (c) exhibits a distinct diagonal structure, highlighting effective melody-to-harmony cross-attention.}
    \label{fig:attn_maps}
\end{figure*}

\section{Harmony and melody representations}

Harmony at the level of chord symbols evolves on a relatively coarse time scale. In the melodic harmonization datasets considered here, quarter-note resolution is sufficient to capture all harmonic detail, since no chords overlap within the same segment. Accordingly, we adopt quarter-note resolution for both melody and harmony, while also comparing results with a finer sixteenth-note resolution.

\textit{Melody} events that occur within each quarter note (or sixteenth note, in the comparison) are grouped and represented as a piano-roll grid. We examine two types of binary matrices: a \textit{full-range} melody roll (fr-roll) $\mathbf{FR} \in \{0,1\}^{L \times P}$ and a \textit{pitch-class} roll (pc-roll) $\mathbf{PC} \in \{0,1\}^{L \times 12}$, where $L$ is the number of time steps and $P=88$ corresponds to MIDI pitches 21 (A0)–108 (C8). The pc-roll encodes chroma information at each timestep, similar to the approach in~\cite{rhyu2022translating}, supporting reasoning over harmonic context. The fr-roll preserves full pitch-range information. We also experiment with a concatenated representation (FRPC), combining both along the feature axis.  

\textit{Harmony} is represented as a sequence of chord tokens from a fixed vocabulary $\mathcal{V}$, denoted $\mathbf{y} \in \mathcal{V}^L$. Chord symbols are normalized following the MIR\_eval~\cite{raffel2014mir_eval} standard (e.g., \texttt{Cmaj7} vs. \texttt{C$^\triangle$}). The vocabulary includes $12 \times 29 = 348$ chord types (12 pitch classes × 29 qualities). Harmony is aligned to the selected grid resolution: if a chord spans multiple steps, it is repeated until its duration ends. For example, a \texttt{C:maj7} spanning two beats occupies two quarter-note grid positions (or eight sixteenth-note positions). Special tokens handle missing or incomplete harmonization: \texttt{<nc>} denotes ``no chord,'' while \texttt{<pad>} fills trailing positions beyond the actual harmonization.  

Previous work~\cite{kaliakatsos2025diffusion} incorporated metric information by prepending a binary vector $\mathbf{g}\in\{0,1\}^{16}$ encoding the time signature (14 bits for the numerator, 2 bits for the denominator). In this setup, the input length equals $1+L+L$, e.g., $129$ tokens for $L=64$ time steps.  

In contrast, we propose embedding metric information directly into melody and harmony without a separate time-signature vector. As shown in Figure~\ref{fig:pianoroll}, special \texttt{<bar>} tokens are inserted into the harmony sequence, while the melody piano-roll gains an additional ``bar'' row indicating bar onsets. This extends the melody matrix by one pitch dimension (e.g., $13 \times L$ for pc-roll) with non-zero values only at barline positions. Both melody and harmony sequences thus integrate bar-level information directly into their representations.  

\begin{figure}[!ht]
\includegraphics[width=0.45\textwidth]{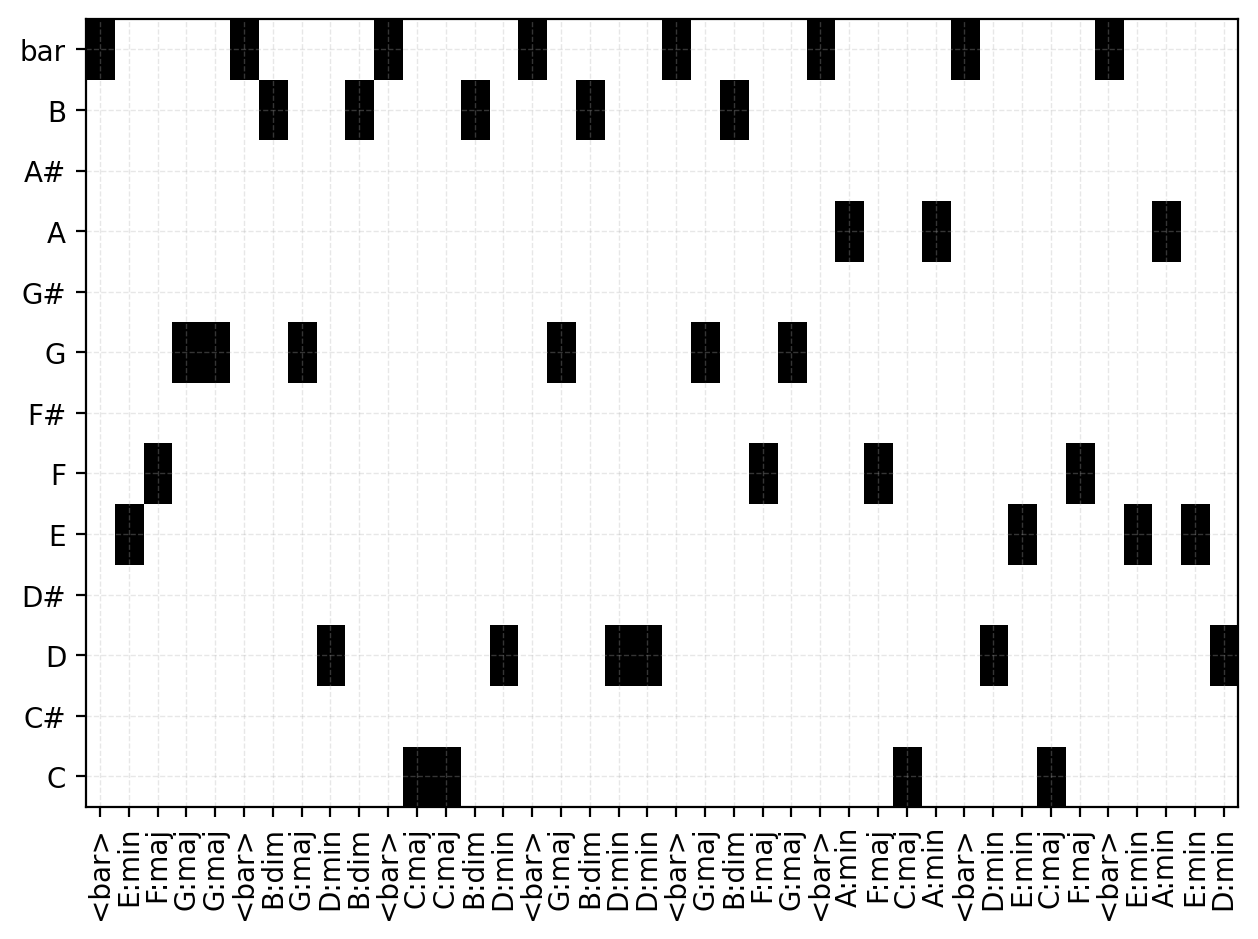}
\caption{Quarter-note resolution of pitch-class pianoroll with integrated bar information. Melody is represented as a $13 \times T$ matrix, with the extra row marking barline positions. Harmony is shown as a parallel sequence of chord tokens with inserted \texttt{<bar>} tokens. A random segment of the artificial dataset described in Section~\ref{subsec:problem} is depicted.}\label{fig:pianoroll}
\end{figure}

\subsection{A training curriculum for promoting cross-attention}

Training in this framework requires strategies for progressively unmasking harmony tokens. Two approaches were proposed in earlier work~\cite{kaliakatsos2025diffusion}, both inspired by diffusion methods:  
\begin{enumerate}
    \item \textbf{Random 10\%} (\texttt{R10}\%) Unmasking: At each stage $k$, 10\% of the remaining masked harmony tokens are randomly revealed, i.e., 10\% unmasking per stage.  
    \item \textbf{Midpoint Doubling} (\texttt{MD}): A deterministic scheme inspired by binary subdivision and musical hierarchy. Tokens are revealed at midpoints between previously visible tokens, doubling the number of unmasked positions at each step.  
\end{enumerate}

In both strategies, training examples are assigned to random stages \textit{regardless of training epoch}. Thus, within a batch, some samples may require harmonization with 0–10\% visibility while others may start with 90–100\%. Importantly, both methods introduce large fractions of visible harmony tokens early in training. As shown in Figure~\ref{fig:attn_maps}(a–b), this encourages the model to rely predominantly on self-attention within the harmony sequence. Since many targets can be reconstructed from local harmonic correlations alone, the model faces little incentive to attend to the melody.  

A likely explanation is that melody supervision is relatively indirect: it propagates through the pianoroll representation projection, whereas harmony–harmony relations are captured directly by the encoder layers. This imbalance places weaker training pressure on cross-attention, particularly in early stages when harmony dominates the visible context. Similar shortcut behavior has been reported in multimodal transformers, where unimodal dependencies override cross-modal cues~\cite{bugliarello2021multimodal, frank2021vision}. Our results suggest that without explicit curriculum design, cross-attention to melody can remain underutilized.  

To address this issue, we propose a curriculum that initially forces the model to rely exclusively on melody. In this method, the number of unmasked tokens \textit{increases with the training epochs} (or epoch subdivision steps). This is the fundamental difference with the existing two methods described above. In the proposed method, during early epochs (or steps), all harmony tokens are masked, compelling the model to establish cross-attention pathways. Harmony tokens are then gradually revealed as the training steps increase, progressing from one visible token up to $L-1$ (where $L$ is the sequence length). This ensures the model learns to handle both extremes: full reliance on melody and partial reliance on harmony.  

The straightforward approach that is proposed is to start training with 0 tokens unmasked (full masking) and gradually move to all but one tokens unmasked (full unmasking). We refer to this strategy as \textbf{full-to-full}. The number of unmasked tokens at each training step is given by:  
\begin{equation}\label{eq:num_unmasked}
    \#\text{unmasked} = \min\left[\ \lfloor v \cdot L \rfloor,\ L-1\ \right],
\end{equation}
where the visible fraction $v$ evolves as:  
\begin{equation}\label{eq:visible_percentage}
    v = \left(\frac{s}{s_\text{total}}\right)^5,
\end{equation}
with $s$ the current training step and $s_\text{total}$ the total number of steps ($0 \leq s < s_\text{total}$). The $\min$ operator ensures at least one masked token remains. The exponent of $5$ allocates roughly half of training to the fully masked regime given typical sequence lengths. This design choice aligns with the goal of forcing early reliance on melody before gradually introducing harmony self-attention. Alternative values (4–10) yielded similar results, so the selected exponent is used as an example value among many possible values. Since our aim is to validate the curriculum concept rather than fine-tune hyperparameters, optimization of this exponent is left to future work.

\subsection{Model architectures}\label{subsec:models}

The core of the presented single encoder transformer architecture is inspired by BERT~\cite{devlin2019bert}, and adapted for generation via masked language modeling (MLM). The model is trained to predict chord tokens conditioned on a melodic context, and varying amounts of visible (i.e., unmasked) portion of the harmony. During inference, the harmony sequence is initially fully masked using \texttt{<mask>} tokens. At each unmasking stage $k$, a subset of partially masked harmony tokens $\mathbf{y}_{\text{in}}^{(k)}$ is provided as part of the overall input. Some portion of those masked tokens needs to be predicted and unmasked, depending on $k$, progressively revealing the complete harmony sequence.

This paper examines the simplest variation of the single encoder generative architecture. During training, the model learns this iterative unmasking process by estimating the conditional distribution over target tokens given the visible context:
\begin{equation}
p_\theta\left( \mathbf{y}_{\text{target}}^{(k)} \mid \mathbf{y}_{\text{in}}^{(k)}, \mathbf{m} \right),
\label{eq:conditional-prob}
\end{equation}
where $\mathbf{y}_{\text{target}}^{(k)}$ contains the subset of harmony tokens to be predicted at stage $k$ and $\mathbf{m}$ denotes melody information either in the pitch-class (PC), or the combined full-range and pitch-class (FRPC) pianorolls. The more complex variations presented in~\cite{kaliakatsos2025diffusion} included also time signature information ($\mathbf{g}$) and the unmaksing step ($t$) explicitly as model inputs:
\begin{equation}
p_\theta\left( \mathbf{y}_{\text{target}}^{(k)} \mid \mathbf{y}_{\text{in}}^{(k)}, \mathbf{m}, \mathbf{g}, k \right),
\label{eq:conditional-prob_old}
\end{equation}
Both formulations enable harmonization in a non-autoregressive manner, while preserving the ability to condition on both the melodic input and previously revealed chords. Another difference between the variations presented in~\cite{kaliakatsos2025diffusion} and the one presented in this paper is that the positional embeddings in our variation are not learnable; instead they are fixed and repeat in the melody and harmony parts of the input. The examined architecture proposed in this paper is kept with fixed positional embeddings for simplification.

The overall input of the transformer layer in the simplified model is formed by passing the melody ($\mathbf{m}$) through a linear projection ($W_\mathbf{m}$) and the current harmony input ($\mathbf{y}_{\text{in}}^{(k)}$) through an embedding ($E_\mathbf{y}$), for matching the dimensionality of the transformer model. The final input of the simplified model is obtained by:
\begin{equation}
\mathbf{z}^{(k)} = \text{concat}\left[ W_\mathbf{m}\left(\mathbf{m} \right), E_\mathbf{y}(\mathbf{y}_{\text{in}}^{(k)}) \right],
\end{equation}
At each step, this overall input embedding is combined with repeating positional embeddings that cover the sequence length $\text{concat}\left[ p,p \right]$ to form the final input to the transformer encoder model as follows:
\begin{equation}
\mathbf{x}^{(k)} = W_{x} \left( \mathbf{z}^{(k)} + \mathbf{\mathit{p}} \right),
\label{eq:embedding-concat}
\end{equation}
where $W_{x}$ is a linear projection layer that maps the position-informed input to the dimensionality of the encoder. The more elaborate versions of~\cite{kaliakatsos2025diffusion} integrate time signature information and stage embeddings with their own projections and embeddings in the overall transformer input -- the reader is referred to this paper for details.

\subsection{Training and inference}\label{subsec:training_inference}

For all training unmasking methods, let $\mathcal{M}^{(k)} \subseteq \{1, \dots, K\}$ denote the set of masked token positions at stage $k$, and $\mathcal{U}^{(k)} \subseteq \mathcal{M}^{(k-1)}$ the set of tokens newly selected for unmasking. Thus, 
\[
\mathcal{M}^{(k-1)} = \mathcal{M}^{(k)} \cup \mathcal{U}^{(k)}.
\]  
In the previously introduced approaches (\texttt{MD}, \texttt{R10}\%), each stage index $k$ is sampled independently of overall training progress, and $\mathcal{U}^{(k)}$ may contain multiple tokens. In the proposed \texttt{FF} strategy, the notion of stage is instead tied to the number of unmasked tokens (Eq.~\ref{eq:num_unmasked}). Here, $\mathcal{U}^{(k)}$ always contains a single randomly chosen token from the still-masked set $\mathcal{H}-\mathcal{M}^{(k)}$, where $\mathcal{H}$ is the set of all harmony tokens.  

The model input at stage $k$ is defined as:
\begin{equation}
y_i^{(k)} =
\begin{cases}
y_i, & i \notin \mathcal{M}^{(k)} \\
\texttt{<mask>}, & i \in \mathcal{M}^{(k)},
\end{cases}
\label{eq:masking-rule}
\end{equation}
and the prediction targets are the unmasked positions:
\begin{equation}
\mathbf{y}_{\text{target}}^{(t)} = \{ y_i \mid i \in (\mathcal{M}^{(k)})^c \},
\label{eq:prediction-targets}
\end{equation}
where $(\mathcal{M}^{(k)})^c = \mathcal{H}-\mathcal{M}^{(k)}$.  

\paragraph{Training}
During \textit{training}, we minimize a masked language modeling loss restricted to the current unmasking step:
\begin{equation}
\mathcal{L}^{(k)} = - \sum_{i \in (\mathcal{M}^{(k)})^c} \log p_\theta(y_i \mid \text{context}),
\label{eq:loss-step}
\end{equation}
with the full loss aggregated over stages:
\begin{equation}
\mathcal{L} = \sum_{k=1}^{K} \mathcal{L}^{(k)},
\label{eq:loss-total}
\end{equation}
where $K$ is the total number of unmasking stages.  
The context includes the melody input $\mathbf{m}$, partially visible harmony $\mathbf{y}_{\text{in}}^{(k)}$, and, depending on the model version, the time signature condition $\mathbf{g}$ and for \texttt{MD} and \texttt{R10}\% the stage condition $k$.

In the \texttt{MD} and \texttt{R10}\% training methods, each batch item is assigned a random stage index $k$, drawn uniformly from $\{1,\dots,K\}$. In contrast, the \texttt{FF} curriculum deterministically increases the number of visible tokens according to Eq.~\ref{eq:num_unmasked}, starting from fully masked harmony. This enforces early reliance on the melody before gradually introducing self-attention from partial harmony input.

\paragraph{Inference}  
Generation begins from a fully masked harmony sequence. We compare three unmasking strategies; the first two are denoted with prefix \texttt{u} to distinguish them from the respective training methods:
\begin{description}
    \item[\texttt{uR10}\%] Selects the top 10\% most confident predictions at each step, compatible with the \texttt{R10}\% unmasking training method.
    \item[\texttt{uMD}] Reveals midpoint tokens deterministically, matching the \texttt{MD} schedule.
    \item[\texttt{Seq}] Sequentially unmasks tokens from left to right, analogous to autoregressive decoding.
\end{description}
It should be noted that the number of model calls for generating a complete harmonic sequence for each strategy is a follows: for \texttt{uR10}\% 10 steps, regardless of sequence length; for \texttt{uMD} $\lceil \text{log}_2(L) \rceil$ steps; for \texttt{Seq} $L$ steps. Therefore, the former two methods are more efficient (for $L > 10$).

Once positions are selected, predictions are drawn from the model distribution:
\begin{equation}
\hat{\mathbf{y}}^{(k)} \sim p_\theta(\cdot \mid \text{context}),
\label{eq:sampling-inference}
\end{equation}
and incorporated into the next input:
\begin{equation}
\mathbf{y}_{\text{in}}^{(k+1)} = \mathbf{y}_{\text{in}}^{(k)} \cup \hat{\mathbf{y}}^{(k)}.
\label{eq:inference-update}
\end{equation}
For all experiments, we used nucleus sampling ($p=0.9$) with temperature $0.2$.  

\paragraph{Implementation details.}  
Models were trained using AdamW with learning rate $1\cdot 10^{-4}$, batch size 8. Each model was trained for 200 epochs with early stopping based on validation loss for the \texttt{MD} and \texttt{R10}\% methods; for \texttt{FF} we saved the model at the last epoch, since it is necessary to keep the model version that has been trained with all the sets of the gradual unmasking curriculum. Training was performed on three NVIDIA RTX3080 GPUs. The loss was averaged over tokens and batches.

\section{Results}\label{sec:results}

This section presents the datasets, evaluation protocols, and experimental findings. We analyze combinations of training curricula, melody representations, tokenization variants, and generation strategies, framing the discussion around five research questions.

\subsection{Dataset and evaluation metrics}\label{subsec:data}

All models are trained on the HookTheory dataset~\cite{yeh2021automatic}, which contains 15,440 MIDI pieces. Following prior work~\cite{rhyu2022translating, huang2024emotion}, we preprocess the data to reflect harmonic rhythm by removing redundant chord repetitions within bars (while allowing carryovers across bar boundaries) and apply key normalization by transposing all major-mode pieces to C major and minor-mode pieces to A minor, using the Krumhansl key-finding algorithm~\cite{krumhansl2001cognitive}. This addresses tonal imbalance and leverages the shared pitch-class structures of C major and A minor~\cite{hahn2024senthymnent}. We split the dataset into 14,679 training and 761 validation/test pieces (95\%/5\%). Evaluation is conducted in two settings~\cite{rhyu2022translating}: (i) \emph{in-domain}, using the HookTheory validation/test split; and (ii) \emph{out-of-domain}, using a curated collection of 650 jazz standard lead sheets, also transposed to C major/A minor. Unlike~\cite{rhyu2022translating}, we do not use the Chord Melody Dataset (CMD)~\cite{chordmelodydataset}, which imposes constraints on chords per bar and note resolution; our curated jazz dataset is also larger (650 vs. 473 pieces). Although the FF curriculum does not introduce bottlenecks that would limit scaling to substantially larger or stylistically diverse datasets, expanding the dataset remains an important direction for future work.

Model outputs are compared against ground-truth harmonizations using music-specific metrics that capture three aspects: chord progression structure, harmony–melody alignment, and rhythmic coherence. For chord progressions, we compute Chord Histogram Entropy (\textbf{CHE}), reflecting the balance of chord usage; Chord Coverage (\textbf{CC}), the number of distinct chord types; and Chord Tonal Distance (\textbf{CTD}), the average tonal distance between adjacent chords, with lower values indicating smoother progressions~\cite{yeh2021automatic, rhyu2022translating, sun2021melody, huang2024emotion}. Harmony–melody alignment is assessed using the Chord Tone to non-Chord Tone Ratio (\textbf{CTnCTR}), measuring how well melody notes align with chord tones; the Pitch Consonance Score (\textbf{PCS}), assigning consonance values to melody–chord intervals; and the Melody–Chord Tonal Distance (\textbf{MCTD}), quantifying tonal proximity between melody and harmony. Finally, rhythmic coherence is analyzed with Harmonic Rhythm Histogram Entropy (\textbf{HRHE}), describing the diversity of chord-change timings; Harmonic Rhythm Coverage (\textbf{HRC}), capturing the variety of rhythmic placement patterns; and Chord Beat Strength (\textbf{CBS}), which evaluates the alignment of chord onsets with metrical accents, where lower scores reflect stronger beat alignment and higher scores reflect syncopation~\cite{wu2024generating}. Table~\ref{tab:ground_truth} reports the average values of these metrics for in-domain and out-of-domain datasets. For all reported metrics, lower mean absolute differences indicate better performance.

\begin{table*}[ht]
  \centering
  \caption{Average metric values for all pieces in the\textit{in-domain} and \textit{out-of-domain} datasets.}
  \label{tab:ground_truth}
  \resizebox{\textwidth}{!}{%
\begin{tabular}{lrrrrrrrrr}
\toprule
Ground truth & CHE & CC & CTD & CTnCTR & PCS & MCTD & HRHE & HRC & CBS \\
\midrule
Test set (in-domain) & 1.4078 & 4.9485 & 0.9748 & 0.7769 & 0.4060 & 1.4139 & 0.4542 & 1.9710 & 0.2314 \\
Jazz set (out-of-domain) & 2.2027 & 11.6471 & 0.8208 & 0.8297 & 0.3145 & 1.4042 & 0.5093 & 2.0607 & 0.2426 \\
\bottomrule
\end{tabular}%
}
\end{table*}

\subsection{Comparisons}

Statistical significance values could indicate if indeed a particular setup is evidently better than any other. The presented values, however, provide indications that the hypotheses entailed by our research questions are not outright rejected. Due to the large number of comparisons that would be required for testing statistical significance, which would necessarily lead to difficulties in presenting such comparisons concisely, statistical significance testing is not performed. The analysis that follows is not intended to show clearly winning setups, but rather indicate what setups appear to be useful.

We compare multiple model setups that vary in architecture, melody and harmony representations, training curricula, and inference strategies. Our evaluation addresses five research questions, applied to both in-domain and out-of-domain datasets:
\begin{description}
    \item[Q1] Does the proposed \texttt{FF} training curriculum outperform existing \texttt{MD} and \texttt{R10}\% approaches?
    \item[Q2] Is quarter-note quantization more effective than sixteenth-note quantization?
    \item[Q3] Is it preferable to encode bar information directly in melody/harmony representations, or to provide time signature as an external condition?
    \item[Q4] Is full-range melody representation necessary, or can pitch-class pianoroll alone achieve strong results (\texttt{FRPC} vs. \texttt{PC})?
    \item[Q5] Does the \texttt{FF} curriculum remain effective when paired with more efficient inference strategies (\texttt{uR}$n$\% and \texttt{uMD})?
\end{description}

Each experimental setup is labeled using the convention:
\begin{verbatim}
    TRAIN_QUANT_BAR_MEL_UNMASK
\end{verbatim}
Setup components are separated with underscores and each setup field includes the following:
\begin{description}[style=unboxed]
    \item[\texttt{TRAIN}]: training curriculum (\texttt{MD}, \texttt{R10}\%, or \texttt{FF}).
    \item[\texttt{QUANT}]: quantization level (\texttt{q4} = quarter-note, \texttt{q16} = sixteenth-note).
    \item[\texttt{BAR}]: bar encoding (\texttt{bar} = intertwined bar info, \texttt{ts} = time signature as condition).
    \item[\texttt{MEL}]: melody representation (\texttt{PC} = pitch-class pianoroll only, \texttt{FRPC} = full-range + pitch class).
    \item[\texttt{UNMASK}]: inference unmasking strategy (\texttt{uR10}\%, \texttt{uMD}, or \texttt{Seq}).
\end{description}

Since~\cite{kaliakatsos2025diffusion} reported that \texttt{MD} outperforms \texttt{R10}\%, we focus primarily on \texttt{MD}-based variants to limit the combinatorial explosion of setups. To measure similarity between generated harmonizations and ground truth, we compute mean absolute differences for each evaluation metric. Results are reported in Table~\ref{tab:in_domain} for the HookTheory in-domain test set and in Table~\ref{tab:out_of_domain} for the out-of-domain jazz dataset.

\begin{table*}[ht]
  \centering
  \caption{Evaluation results in the\textit{in-domain} test dataset. Mean absolute difference differences are calculated, and the smallest differences per metric are show in bold.}
  \label{tab:in_domain}
  \resizebox{\textwidth}{!}{%
\begin{tabular}{lrrrrrrrrr}
\toprule
Model & CHE & CC & CTD & CTnCTR & PCS & MCTD & HRHE & HRC & CBS \\
\midrule
\texttt{R10}\%\_\texttt{q4}\_\texttt{bar}\_\texttt{PC}\_\texttt{Seq} & 0.7311 & 2.7414 & 0.4956 & 0.1303 & 0.1420 & 0.1170 & 0.4543 & \textbf{0.9749} & 0.2307 \\
\texttt{R10}\%\_\texttt{q4}\_\texttt{bar}\_\texttt{PC}\_\texttt{uR10}\% & 0.7306 & 2.6873 & 0.5111 & 0.1310 & 0.1441 & 0.1140 & 0.4643 & 1.0092 & 0.2274 \\ \hline
\texttt{MD}\_\texttt{q16}\_\texttt{ts}\_\texttt{FRPC}\_\texttt{uMD} & 0.7301 & 2.4934 & 0.4862 & 0.1282 & 0.1471 & 0.1157 & 0.5505 & 1.4446 & 0.2711 \\
\texttt{MD}\_\texttt{q16}\_\texttt{bar}\_\texttt{FRPC}\_\texttt{uMD} & 0.5941 & 2.2111 & 0.4371 & 0.1145 & 0.1223 & 0.0933 & 0.4599 & 1.0079 & 0.2223 \\
\texttt{MD}\_\texttt{q4}\_\texttt{ts}\_\texttt{FRPC}\_\texttt{uMD} & 0.9177 & 2.9446 & 0.6737 & 0.1321 & 0.1390 & 0.1085 & 0.5383 & 1.2995 & 0.2286 \\
\texttt{MD}\_\texttt{q4}\_\texttt{bar}\_\texttt{FRPC}\_\texttt{uMD} & 1.1031 & 3.3661 & 0.7320 & 0.1398 & 0.1437 & 0.1147 & 0.5374 & 1.2445 & 0.2355 \\
\texttt{MD}\_\texttt{q4}\_\texttt{bar}\_\texttt{PC}\_\texttt{uMD} & 1.2849 & 3.6889 & 0.8927 & 0.1433 & 0.1434 & 0.1160 & 0.4949 & 1.1304 & 0.2390 \\ \hline
\texttt{FF}\_\texttt{q16}\_\texttt{ts}\_\texttt{FRPC}\_\texttt{Seq} & 0.4908 & 2.3509 & 0.4063 & 0.1502 & 0.1657 & 0.1226 & 0.8482 & 3.1860 & 0.3633 \\
\texttt{FF}\_\texttt{q16}\_\texttt{ts}\_\texttt{PC}\_\texttt{Seq} & 0.9386 & 2.7704 & 0.7184 & 0.2044 & 0.2297 & 0.1832 & 0.5933 & 1.9855 & 0.2974 \\
\texttt{FF}\_\texttt{q16}\_\texttt{bar}\_\texttt{FRPC}\_\texttt{Seq} & 0.5228 & 2.0699 & 0.4517 & 0.1366 & 0.1388 & 0.1126 & 0.4597 & 1.1385 & 0.2300 \\
\texttt{FF}\_\texttt{q16}\_\texttt{bar}\_\texttt{PC}\_\texttt{Seq} & 0.6862 & 2.3166 & 0.5130 & 0.1580 & 0.1811 & 0.1282 & 0.4959 & 1.2889 & 0.2300 \\
\texttt{FF}\_\texttt{q4}\_\texttt{ts}\_\texttt{FRPC}\_\texttt{Seq} & 0.4467 & 2.2243 & 0.3334 & 0.1130 & 0.1152 & \textbf{0.0864} & 0.4688 & 1.1425 & 0.1988 \\
\texttt{FF}\_\texttt{q4}\_\texttt{ts}\_\texttt{PC}\_\texttt{Seq} & \textbf{0.3892} & \textbf{1.8377} & 0.3092 & 0.1170 & 0.1241 & 0.0932 & 0.4662 & 1.1108 & 0.1916 \\
\texttt{FF}\_\texttt{q4}\_\texttt{bar}\_\texttt{PC}\_\texttt{uMD} & 0.4328 & 2.4749 & \textbf{0.2660} & \textbf{0.1086} & \textbf{0.1132} & 0.0871 & 0.5268 & 1.3259 & \textbf{0.1858} \\
\texttt{FF}\_\texttt{q4}\_\texttt{bar}\_\texttt{PC}\_\texttt{Seq} & 0.4237 & 2.1280 & 0.2816 & 0.1091 & 0.1176 & 0.0909 & \textbf{0.4422} & 1.0422 & 0.1888 \\
\texttt{FF}\_\texttt{q4}\_\texttt{bar}\_\texttt{PC}\_\texttt{uR10}\% & 0.4030 & 2.0963 & 0.2766 & 0.1091 & 0.1136 & 0.0890 & 0.5211 & 1.2902 & 0.1910 \\
\bottomrule
\end{tabular}%
}
\end{table*}

\begin{table*}[h!]
  \centering
  \caption{Evaluation results in the\textit{out-of-domain} test dataset. Mean absolute difference differences are calculated, and the smallest differences per metric are show in bold.}
  \label{tab:out_of_domain}
  \resizebox{\textwidth}{!}{%
\begin{tabular}{lrrrrrrrrr}
\toprule
Model & CHE & CC & CTD & CTnCTR & PCS & MCTD & HRHE & HRC & CBS \\
\midrule
\texttt{R10}\%\_\texttt{q4}\_\texttt{bar}\_\texttt{PC}\_\texttt{Seq} & 1.3892 & 9.0436 & 0.4780 & 0.2203 & 0.1178 & 0.1656 & 0.5060 & 1.0550 & 0.2421 \\
\texttt{R10}\%\_\texttt{q4}\_\texttt{bar}\_\texttt{PC}\_\texttt{uR10}\% & 1.4190 & 9.0417 & 0.4451 & 0.2250 & 0.1219 & 0.1682 & 0.4880 & \textbf{1.0417} & 0.2384 \\ \hline
\texttt{MD}\_\texttt{q16}\_\texttt{ts}\_\texttt{FRPC}\_\texttt{uMD} & 1.4335 & 8.7761 & 0.3958 & 0.1762 & 0.1157 & 0.1227 & 0.4120 & 1.2410 & 0.1989 \\
\texttt{MD}\_\texttt{q16}\_\texttt{bar}\_\texttt{FRPC}\_\texttt{uMD} & 1.0667 & 7.6395 & 0.3125 & 0.1227 & 0.1195 & 0.0844 & 0.4955 & 1.1480 & 0.2366 \\
\texttt{MD}\_\texttt{q4}\_\texttt{ts}\_\texttt{FRPC}\_\texttt{uMD} & 1.2569 & 8.3055 & 0.3918 & 0.1604 & 0.1019 & 0.1091 & 0.4635 & 1.3700 & 0.1912 \\
\texttt{MD}\_\texttt{q4}\_\texttt{bar}\_\texttt{FRPC}\_\texttt{uMD} & 2.0603 & 10.4469 & 0.7242 & 0.1877 & 0.1147 & 0.1291 & 0.5592 & 1.2325 & 0.2481 \\
\texttt{MD}\_\texttt{q4}\_\texttt{bar}\_\texttt{PC}\_\texttt{uMD} & 2.1433 & 10.5960 & 0.7820 & 0.1959 & 0.1050 & 0.1288 & 0.5449 & 1.1600 & 0.2573 \\ \hline
\texttt{FF}\_\texttt{q16}\_\texttt{ts}\_\texttt{FRPC}\_\texttt{Seq} & 0.7635 & 5.1992 & 0.2242 & 0.2160 & 0.1104 & 0.1429 & 0.9303 & 4.5123 & 0.3405 \\
\texttt{FF}\_\texttt{q16}\_\texttt{ts}\_\texttt{PC}\_\texttt{Seq} & 1.2961 & 7.7438 & 0.3851 & 0.2277 & 0.1413 & 0.1672 & 0.7017 & 3.6641 & 0.3016 \\
\texttt{FF}\_\texttt{q16}\_\texttt{bar}\_\texttt{FRPC}\_\texttt{Seq} & 0.7240 & 5.4972 & 0.2374 & 0.1728 & 0.1053 & 0.1198 & 0.4214 & 1.5161 & 0.2145 \\
\texttt{FF}\_\texttt{q16}\_\texttt{bar}\_\texttt{PC}\_\texttt{Seq} & 1.0161 & 6.3757 & 0.3565 & 0.1814 & 0.1196 & 0.1248 & 0.4536 & 1.7989 & 0.2220 \\
\texttt{FF}\_\texttt{q4}\_\texttt{ts}\_\texttt{FRPC}\_\texttt{Seq} & 0.4577 & 4.0702 & 0.2082 & 0.1062 & 0.1046 & 0.0801 & 0.4027 & 1.2543 & 0.1725 \\
\texttt{FF}\_\texttt{q4}\_\texttt{ts}\_\texttt{PC}\_\texttt{Seq} & 0.4984 & 4.2732 & 0.2047 & 0.1082 & \textbf{0.0992} & 0.0820 & 0.4027 & 1.2505 & 0.1734 \\
\texttt{FF}\_\texttt{q4}\_\texttt{bar}\_\texttt{PC}\_\texttt{uMD} & \textbf{0.3854} & \textbf{3.6641} & 0.2064 & \textbf{0.0895} & 0.1096 & 0.0771 & 0.4622 & 1.4118 & \textbf{0.1615} \\
\texttt{FF}\_\texttt{q4}\_\texttt{bar}\_\texttt{PC}\_\texttt{Seq} & 0.4557 & 3.9943 & 0.2167 & 0.1019 & 0.1113 & 0.0789 & \textbf{0.3675} & 1.1195 & 0.1781 \\
\texttt{FF}\_\texttt{q4}\_\texttt{bar}\_\texttt{PC}\_\texttt{uR10}\% & 0.4567 & 3.9886 & \textbf{0.1936} & 0.0903 & 0.1088 & \textbf{0.0754} & 0.3919 & 1.2448 & 0.1622 \\
\bottomrule
\end{tabular}%
}
\end{table*}

Tables~\ref{tab:in_domain} and \ref{tab:out_of_domain} show that the \texttt{FF} training protocol produces better results for both in- and out-of-domain sets under all examined metrics, except HRC, where the \texttt{R10}\% variant performs slightly better. Notably, for the CHE and CC metrics -- which capture the most pronounced differences between the in-domain and out-of-domain datasets as shown in Table~\ref{tab:ground_truth} -- \texttt{FF}-trained models achieve substantially lower mean absolute errors compared to other training protocols. This suggests that \texttt{FF}-trained models are better able to select chords appropriate for the jazz style. Given that all models are conditioned only on the melody, the superior performance of the \texttt{FF} variants on the jazz (out-of-domain) set under the CHE and CC metrics likely reflects their capacity to leverage melodic nuances more effectively by employing a wider variety of chord symbols. In contrast, the \texttt{MD} and \texttt{R10}\% models appear to rely more rigidly on their learned harmonic patterns. The answer to \textbf{Q1} is therefore that the \texttt{FF}-trained models consistently outperform the other training curricula.

Regarding the second question, Tables~\ref{tab:in_domain} and \ref{tab:out_of_domain} show that \texttt{q4} resolution is consistently present in all top-performing model variants. While the \texttt{FF} variants perform slightly better under \texttt{q4} resolution across nearly all metrics, this pattern does not hold as clearly for the \texttt{MD} variants, where some metrics favor \texttt{q16}. Thus, the answer to \textbf{Q2} is that \texttt{q4} resolution provides a clear advantage for the best-performing \texttt{FF}-trained models.

With respect to bar intertwining vs. time signature conditioning, incorporating bar information in the tokenization led to slightly more winning combinations for the \texttt{FF} variants in the in-domain dataset (4 \texttt{bar} vs. 3 \texttt{ts} cases; Table~\ref{tab:in_domain}). The advantage becomes more pronounced in the out-of-domain dataset (Table~\ref{tab:out_of_domain}), where nearly all metrics favored \texttt{bar} over \texttt{ts}. For \texttt{MD} variants the pattern is less consistent, though the \texttt{q16}-\texttt{bar} combination performs relatively well. Therefore, the answer to \textbf{Q3} is that bar intertwining generally yields better results for \texttt{FF}-trained models.

On the question of melody representation, results show that using only pitch classes (\texttt{PC}) is not only sufficient but in fact advantageous. In Table~\ref{tab:in_domain}, only one metric marginally favors \texttt{FRPC}, while in Table~\ref{tab:out_of_domain}, all best-performing \texttt{FF} variants rely solely on \texttt{PC}. For the \texttt{MD} variants the results are again less consistent. Thus, the answer to \textbf{Q4} is that pitch class–only pianoroll representations are generally superior for \texttt{FF}-trained models.

Finally, when comparing unmasking strategies at inference time, the best-performing \texttt{FF}-trained models are distributed across all three strategies (\texttt{uMD}, \texttt{uR10}\%, and \texttt{Seq}), with no single method dominating. The key observation is that both \texttt{uMD} and \texttt{uR10}\% appear among the top-performing setups, which is notable because they require fewer model calls to generate complete harmonizations (see Section~\ref{subsec:training_inference}). The answer to \textbf{Q5} is that the \texttt{FF} training curriculum remains effective when paired with more efficient unmasking strategies.

\section{Conclusions}\label{sec:conclusions}

This paper addressed the problem of weak melody (cross) attention between in previously proposed single-encoder harmonization methods. Prior approaches based on diffusion-inspired unmasking training strategies tended to underutilize melodic information, as evidenced by diffuse or absent attention patterns. To overcome this limitation, we introduced the \texttt{FF} (full-to-full) training curriculum, which keeps all-mask harmony tokens during the initial training steps and progressively unmasks the full sequence of tokens. In combination with different quantizations, structure-related tokenization schemes, melody representations, and inference strategies, we evaluated the effectiveness of this approach across both in-domain and out-of-domain datasets.

Our experimental analysis answered a set of targeted research questions. First, the proposed \texttt{FF} training curriculum clearly outperformed the existing \texttt{MD} and \texttt{R10}\% protocols in nearly all metrics, especially in out-of-domain settings where it proved more capable of adapting harmonic choices to the underlying melody (Q1). Second, quarter-note quantization (\texttt{q4}) consistently yielded stronger results than sixteenth-note quantization (\texttt{q16}) for the \texttt{FF} models (Q2). Third, intertwining bar information in the tokenization generally improved performance compared to using time signatures as conditions (Q3). Fourth, representing melodies with only pitch classes (\texttt{PC}) was sufficient and in most cases superior to using full-range pitch representations (\texttt{FRPC}) (Q4). Finally, the \texttt{FF} curriculum maintained its effectiveness across all inference-time unmasking strategies, including the more efficient \texttt{uMD} and \texttt{uR10}\% protocols (Q5). The single-encoder architecture used here is intentionally minimal, serving as a controlled environment for isolating the effect of the curriculum itself. Future work will extend the method to richer and more expressive model variants. 

For future research, an interesting direction would be to investigate the proportion of masked tokens during training at each individual step. The exponent in Equation~\ref{eq:visible_percentage} has been shown to yield good results within the explored range of values for the examined dataset. However, further improvements might be achieved by dynamically monitoring the behavior of cross and self attention during training and adapting the percentage of visible harmony tokens accordingly. In future extensions, we also plan to supplement quantitative metrics with qualitative listening studies and curated harmonization examples, enabling a more perceptual assessment of the FF curriculum’s impact. Moreover, different harmonization styles --or even individual pieces within a batch-- may vary in the degree to which harmonic structure depends on stylistic conventions versus melodic context. It would therefore be valuable to explore whether such interactions can be identified and quantified to compute optimal, piece- or style-specific visibility schedules for harmony tokens during training.

% Looking ahead, several avenues remain open for exploration. Future work could investigate integrating explicit style-conditioning signals to improve generalization across genres, or extending the framework to multi-part harmonization beyond chord symbols. Incorporating external harmonic priors, such as probabilistic transition models or music-theoretic rules, could further enhance controllability during generation. Finally, adapting the full-to-full training strategy to other sequence generation domains—such as symbolic rhythm generation, polyphonic texture completion, or even beyond music—may reveal its broader applicability as a general-purpose unmasking curriculum.

\section*{Acknowledgment}
This work has been partially supported by project MIS 5154714 of the National Recovery and Resilience
Plan Greece 2.0 funded by the European Union under the NextGenerationEU Program.

\bibliographystyle{IEEEtran}
\bibliography{references}
% \bibitem{b1} G. Eason, B. Noble, and I. N. Sneddon, ``On certain integrals of Lipschitz-Hankel type involving products of Bessel functions,'' Phil. Trans. Roy. Soc. London, vol. A247, pp. 529--551, April 1955.
% \bibitem{b2} J. Clerk Maxwell, A Treatise on Electricity and Magnetism, 3rd ed., vol. 2. Oxford: Clarendon, 1892, pp.68--73.
% \bibitem{b3} I. S. Jacobs and C. P. Bean, ``Fine particles, thin films and exchange anisotropy,'' in Magnetism, vol. III, G. T. Rado and H. Suhl, Eds. New York: Academic, 1963, pp. 271--350.
% \bibitem{b4} K. Elissa, ``Title of paper if known,'' unpublished.
% \bibitem{b5} R. Nicole, ``Title of paper with only first word capitalized,'' J. Name Stand. Abbrev., in press.
% \bibitem{b6} Y. Yorozu, M. Hirano, K. Oka, and Y. Tagawa, ``Electron spectroscopy studies on magneto-optical media and plastic substrate interface,'' IEEE Transl. J. Magn. Japan, vol. 2, pp. 740--741, August 1987 [Digests 9th Annual Conf. Magnetics Japan, p. 301, 1982].
% \bibitem{b7} M. Young, The Technical Writer's Handbook. Mill Valley, CA: University Science, 1989.
% \end{thebibliography}
\vspace{12pt}
% \color{red}
% IEEE conference templates contain guidance text for composing and formatting conference papers. Please ensure that all template text is removed from your conference paper prior to submission to the conference. Failure to remove the template text from your paper may result in your paper not being published.

\end{document}